\documentclass[twocolumn,prl,aps,showpacs]{revtex4}
\usepackage{pictex}

\newfont{\bigsym}{cmr10  scaled \magstep2}
\newfont{\dlsbig}{cmmi10 scaled \magstep2}
\newfont{\dlsmed}{cmmi10 scaled \magstep1}

\begin{document}
\title{Exact Free Energy Functional for a Driven Diffusive Open Stationary 
  Nonequilibrium  System}

\author{B. Derrida}
\email{derrida@lps.ens.fr}
\affiliation{Laboratoire de Physique Statistique,
 24 rue Lhomond, 75231 Paris Cedex 05,
 France}

\author{J.~L.~Lebowitz}
\altaffiliation{Institute for Advanced Study, Princeton, NJ 08540 and
Department of Physics, Rutgers}
\email{lebowitz@math.rutgers.edu}

\author{E.~R.~Speer}
\email{speer@math.rutgers.edu}
\affiliation{Department of Mathematics, Rutgers University, 
 New Brunswick, NJ 08903}

\begin{abstract}
 We obtain the exact probability $\exp[-L {\cal F}(\{\rho(x)\})]$ of
finding a macroscopic density profile $\rho(x)$ in the stationary
nonequilibrium state of an open driven diffusive system, when the size of
the system $L \to \infty$.  $\cal F$, which plays the role of a
nonequilibrium free energy, has a very different structure from that found
in the purely diffusive case.  As there, $\cal F$ is nonlocal, but the
shocks and dynamic phase transitions of the driven system are reflected in
non-convexity of $\cal F$, in discontinuities in its second derivatives,
and in non-Gaussian fluctuations in the steady state.
\end{abstract}

\maketitle

% body of paper here

The behavior of macroscopic systems which carry steady currents is one of
the central problems in nonequilibrium statistical mechanics \cite{Ruelle}.
Of particular interest are stationary nonequilibrium states (SNS)
maintained by contact with infinite reservoirs at the system boundaries and
subjected to bulk driving forces.  A paradigm of such systems is a fluid in
contact with a thermal reservoir at temperature $T_a$ at the top and
$T_b>T_a$ at the bottom, for which gravity supplies the bulk force (the
Rayleigh-B\'enard system \cite{CH}); the system exhibits dynamic phase
transitions corresponding to the formation of different patterns of heat
and mass flow as the parameters are varied.  By contrast, if $T_a>T_b$ the
system has no instabilities.  These dynamic transitions are not understood,
at present, in terms of a microscopically derived free energy, despite
various promising attempts \cite{ST}.  Here we obtain the analogue of such
a free energy for the SNS of a model system which, despite its simplicity,
has some dynamic transitions.

We consider the one-dimensional asymmetric simple exclusion process (ASEP)
on a lattice of $L$ sites \cite{krug,DEHP}.  Each site $i$, $i=1,\ldots,L$,
is either occupied by a single particle ($\tau_i=1$) or is empty
($\tau_i=0$).  In the interior of the system ($2 \leq i \leq L-1$), a
particle attempts to jump to its right neighboring site with rate 1 and to
its left neighboring site with rate $q$ (with $0 \leq q < 1$), succeeding
if the target site is empty.  At the boundary site $i=1$ ($i=L$) particles
jump only to the right (left).  These boundary sites are also connected to
particle reservoirs: if site 1 is empty, it becomes occupied at rate
$\alpha$ (by a particle from the left reservoir); similarly, if site $L$ is
occupied, the particle may jump into the right reservoir at rate $\beta$.

This dynamics produces an SNS for which we calculate the large $L$ behavior
of $P_L(\{\rho(x\}) \simeq \exp [-L {\cal F}(\{\rho(x)\})]$, the
probability for observing microscopic configurations corresponding, in the
limit $L\rightarrow\infty$, $x=i/L$, to the macroscopic density profile
$\rho(x)$, $0 \leq x \leq 1$.  ${\cal F}(\{\rho(x)\})$ is generally called
the large deviation functional (LDF) in the mathematical literature
\cite{O-E}; one always has ${\cal F}(\{\rho(x)\})\ge0$, with equality
holding only if $\rho(x)=\bar \rho(x)$, a typical density profile in the
system, so that atypical profiles are observed with exponentially small
probability for large $L$.  For equilibrium systems ${\cal F}(\{\rho(x)\})$
is given as the difference in free energies for $\rho(x)$ and for
$\bar\rho(x)$.

In our previous work \cite{DLSLD} we considered the case $q=1$, in which
the bulk dynamics are symmetric.  We obtained there an exact
${\cal F}_s(\{\rho\})$ which, unlike the free energy in equilibrium, was
nonlocal, reflecting the generic presence of long range correlations in SNS
\cite{HS}.  Due to the purely diffusive nature of the bulk dynamics,
${\cal F}_s$ did not exhibit any phase transitions or instabilities.  This
is quite different from the asymmetric case considered here which has, due
to the driven nature of the bulk dynamics, not only long range correlations
but also a rich phase diagram including phase transitions and shocks
\cite{krug,DEHP,PS}.  It is thus closer to a real fluid and gives rise to a
correspondingly more complex ${\cal F}(\{\rho\})$.

Before describing our new results about $\cal F$ we summarize some known
properties of the open ASEP \cite{Ligg2}.  We restrict ourselves in this
paper to $\alpha = (1-q)\rho_a$ and $\beta = (1-q)(1-\rho_b)$ with
$0 \leq \rho_a, \rho_b \leq 1$; the parameters $\rho_a$ and $\rho_b$
represent the densities in the left and right reservoirs.  For
$\rho_a=\rho_b=r$ the stationary measure is just a product measure
\cite{DEHP,Ligg2} with uniform density $r$.  This means that all static
(i.e.\ single time) properties of the system, including $\cal F$, are the
same as for an equilibrium lattice gas with fugacity $z = r/(1-r)$.  For
this system the LDF is given by \cite{O-E,DLSLD}
 \begin{eqnarray}
  \label{Feq}
 {\cal F}_{eq}(\{\rho(x)\})&& \\
  &&\hskip-50pt
   =\;\int_0^1 \left[\rho(x) \log{\rho(x) \over r}
    +  (1 - \rho(x)) \log{1 - \rho(x) \over 1 - r}\right]\,dx.
   \nonumber
 \end{eqnarray}

 Now recall that if an infinite system with ASEP dynamics is started in an
initial state with a macroscopic density profile $\rho(x,0) = \rho_a$ for
$x < X$, $\rho(x,0) = \rho_b$ for $x > X$, then when $\rho_a < \rho_b$ the
time evolved $\rho(x,t)$ will maintain a sharp shock which will move with
velocity $V = (1-q)(1-\rho_a-\rho_b)$, while when $\rho_a > \rho_b$,
$\rho(x,t)$ will smooth out as in a ``fan'':
 $\rho(x,t)=\rho_a$ if $x\le x_a(t)$,
 $\rho(x,t)=\rho_a+(\rho_b-\rho_a)(x-x_a(t))/(x_b(t)-x_a(t))$
 if $x_a(t)<x<x_b(t)$, and $\rho(x,t)=\rho_b$ if $x_b(t)\le x$, with
 $x_\alpha(t) =X+(1-q)(1-2\rho_\alpha)t$, $\alpha=a,b$.
 In each case (unless $V=0$) the system will attain, as $t \to \infty$, a
constant density $\bar \rho$ in any finite region.  This gives an
understanding \cite{PS} of the phase diagram shown in Fig.~1 for the system
with open boundaries.  For $\rho_a < \rho_b$, the shock will move to the
right boundary when $V > 0$, leaving behind the constant density
$\bar\rho=\rho_a$ (phase $A_1$), and to the left boundary when $V < 0$,
yielding $\bar\rho=\rho_b$ (phase $B_1$).  On the line $S$ ($V=0$) a
typical $\bar \rho(x)$ is no longer constant, but corresponds to a shock at
some point $s$, uniformly distributed on $[0,1]$, where $\bar \rho(x)$
jumps from $\rho_a$ to $\rho_b$, i.e.\
$\bar \rho(x) = \rho_s(x)\equiv\rho_a \Theta(s-x) + \rho_b \Theta(x-s)$
with $\Theta$ the Heaviside function \cite{Ligg2}.  This line corresponds
to a first order phase transition, with $\bar\rho$ discontinuous across
$S$.  For $\rho_a>\rho_b$ (phases $A_2$, $B_2$, and $C$) one sees the
constant density, $\rho_a$, $\rho_b$, or $1/2$, which would have resulted
from the fanlike behavior in the infinite system.

\begin{figure}
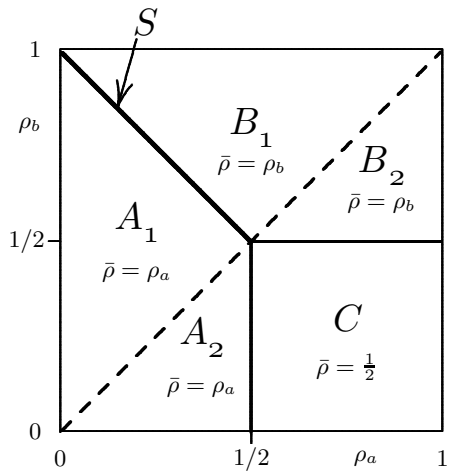

\hbox to \hsize{\hss
 \beginpicture
 \setcoordinatesystem units <2.0truein,2.0truein> point at 0 0
 \setplotarea x from 0.0 to  1.0, y from 0.0 to  1.0
 \plotsymbolspacing=0.1pt
 \setlinear
 \plot  0 0 1 0 1 1 0 1 0 0 /
 \plot 0.5 0.0 0.5 -0.03 /
 \put {1/2}[t] at 0.5 -0.04
 \plot 0.0 0.5 -0.03 0.5  /
 \put {1/2}[r] at -0.04 0.5 
 \arrow <10pt> [0.2,0.6] from 0.2 1.03 to 0.15 0.85
 \put {\dlsbig S} [b] at 0.22 1.04
 \setplotsymbol ({.})
 \setdashes
 \plot 0.004 0.004 0.996 0.996 /
 \setsolid
 \plot 0.998 0.5 0.5 0.5 0.5 0.002 /
 \setplotsymbol ({\bigsym .})
 \plot 0.005 0.995 0.5 0.5 /
 \put {\dlsbig A\lower5pt\hbox{\dlsmed1}} at 0.2 0.55
 \put {$\bar\rho=\rho_a$} at 0.2 0.42
 \put {{\dlsbig A\lower5pt\hbox{\dlsmed2}}} at 0.37 0.25
 \put {$\bar\rho=\rho_a$} at 0.37 0.12
 \put {\dlsbig B\lower5pt\hbox{\dlsmed1}} at 0.5 0.8
 \put {$\bar\rho=\rho_b$} at 0.5 0.70
 \put {{\dlsbig B\hskip1pt\lower5pt\hbox{\dlsmed2}}} at 0.84 0.7
 \put {$\bar\rho=\rho_b$} at 0.84 0.60
 \put {\dlsbig C} at 0.75 0.3
 \put {$\bar\rho={1\over2}$} at 0.75 0.17
 \put {0} [t] at 0 -0.05
 \put {1} [t] at 1 -0.05
 \put {0} [r] at -0.05 0
 \put {1} [r] at -0.05 1
 \put {$\rho_a$} [t] at 0.8 -0.05
 \put {$\rho_b$} [r] at -0.05 0.8
 \endpicture
 \hss}
 \bigskip
 
\caption{The phase diagram of the open ASEP\label{fig1}}

\end{figure}

 We now turn to our new results.  The line $\rho_a = \rho_b$, which
separates what we will call the shock region $\rho_a < \rho_b$ from the fan
region $\rho_a> \rho_b$, is irrelevant for $\bar \rho$ but plays a crucial
role in the LDF.  Defining
 \begin{eqnarray}
 &g(h,f) = h\log[h(1-f)]+(1-h)\log[(1-h)f],& \label{gdef}\\
 &K(\rho_a,\rho_b) = \log \bar\rho(1- \bar\rho),&\label{Kdef}
 \end{eqnarray}
 with $\bar \rho$ given in Fig.\ 1, we find: for $\rho_a > \rho_b$,
 \begin{eqnarray}
 \label{result1}
{\cal F} (\{\rho(x)\};\rho_a,\rho_b) \hskip-40pt&& \\
   &=& - K(\rho_a,\rho_b)   
   +\;\sup_{F(x)} \int_0^1 dx\,g(\rho(x),F(x)), \nonumber
\end{eqnarray}
 where the sup is over all {\it monotone} nonincreasing functions $F(x)$
which satisfy $\rho_a\ge F(x)\ge\rho_b$, $0 \leq x \leq 1$; for
$\rho_a < \rho_b$,
\begin{eqnarray}
 \label{result2}
 {\cal F}  (\{ \rho(x) \};\rho_a,\rho_b)) 
  &=& - K(\rho_a,\rho_b) \\
 &&\hskip-90pt\nonumber
   +\;   \inf_{0 \leq y \leq 1}\left\{\int_0^y dx\,g(\rho(x),\rho_a)
   + \int_y^1 dx\,g(\rho(x),\rho_b)\right\}.\nonumber
\end{eqnarray}

The fact that the function $F(x)$ in (\ref{result1}) is required to be
monotone makes it, like $y$ in (\ref{result2}), depend on the global form
of $\rho(x)$, so ${\cal F}$ is a nonlocal functional of $\rho(x)$.  It can
be shown \cite{DLSALD} that the optimal $F$ in (\ref{result1}), which we
denote by $F_\rho$, is constructed as follows: let $G_\rho$ be defined as
the concave envelope of the function $\int_0^x (1 - \rho(y))\,dy$; then
$G_\rho^\prime(x)$ is monotone,
 \begin{equation} \label{CE}
   F_\rho(x) = G_\rho'(x) \quad {\rm if} \quad  \rho_b \leq G_\rho'(x) \leq \rho_a,
 \end{equation}
 and $F_\rho(x) = \rho_a$ ($F_\rho(x) = \rho_b$) where
$G_\rho'(x)\ge\rho_a$ ($G_\rho'(x)\le\rho_b$).  Note that $F_\rho(x)$ need
not be continuous; it will generally consist of strictly decreasing pieces,
where $F_\rho(x) = 1 - \rho(x)$, flat pieces, where $F_\rho(x)=\rho_a$,
$F_\rho(x)=\rho_b$, or $F_\rho(x)$ is obtained from $1 - \rho(x) $ by a
Maxwell construction (the integrals of $F_\rho(x)$ and of $1 - \rho(x)$
over the latter intervals being equal), and possible jumps downward.

Before going on to discuss the derivation of (\ref{result1}) and
(\ref{result2}) we describe some consequences. 

(a) It can be verified that ${\cal F}(\{\rho(x)\}; \rho_a, \rho_b) \geq 0$;
equality occurs only when $\rho(x) = \bar \rho$ as given in the phase
diagram, {\it except} on the first order line $S$, where it is the shock
(typical) configurations $\rho_s(x)$ which satisfy
${\cal F}(\{\rho_s(x)\}; \rho_a, \rho_b)= 0$ for all $s\in[0,1]$.

(b) ${\cal F}(\{\rho(x)\}; \rho_a,\rho_b)$ given by (\ref{result1}) is a
convex functional of $\rho(x)$ in the fan region $\rho_a \geq \rho_b$,
since it is the supremum over $F$ of $\int_0^1 g(\rho,F)dx$ with $g$ a
locally convex function of $\rho$ for every $F$.  This is similar to what
happens in the symmetric case \cite{DLSLD}.  In the shock region
$\rho_a < \rho_b$, on the contrary, this is not true; for every
$\rho_a,\rho_b$ there are profiles $\rho(x)$ near which $\cal F$ is not
convex. This is easily verified on the line $S$ where a superposition
$\rho(x)=\lambda\rho_s(x)+(1-\lambda)\rho_t(x)$, $s\ne t$, satisfies
${\cal F}(\{\rho(x)\})>0$ for $0<\lambda<1$.

(c) The LDF in the fan region $\rho_a>\rho_b$ has similarities besides
convexity to that in the symmetric case.  In particular it is easy to see
from (\ref{result1}) that
${\cal F}(\{\rho(x)\}) \geq {\cal F}_{eq}(\{\rho(x)\})$, where
${\cal F}_{eq}$ is given by (1) with $r$ replaced by the appropriate
$\bar \rho$.  But in the shock region, $\rho_a < \rho_b$, this inequality
is reversed: ${\cal F}(\{\rho(x)\}) \leq {\cal F}_{eq}(\{\rho(x)\})$ (as is
clear from (\ref{result2}), since in region $B_1$ ($A_1$), $y=0$ ($y=1$)
gives ${\cal F}_{\rm eq}(\{\rho(x)\})$).  This is similar to what
fluctuating hydrodynamics predicts for the Rayleigh-B\'enard problem
discussed earlier: fluctuations are decreased when $T_a>T_b$ (at least when
$T_a-T_b$ is very small) and are increased when $T_a<T_b$, even in the
stable conductance regime \cite{fluct_Ben}.

(d) In the symmetric case discussed in \cite{DLSALD}, as in an equilibrium
system not at a phase transition (in any dimension), the probability of
small fluctuations about the typical state can be obtained from ${\cal F}$
as a limit.  More precisely if we write
$\rho(x) = \bar \rho(x) + {1 \over \sqrt L} u(x)$ and then expand $\cal F$
to second order (the first order term being zero) we get a Gaussian
distribution for $u(x)$ with covariance $C(x,x')$, where
$C^{-1}(x,x') = \delta^2 {\cal F}/\delta \rho(x) \delta \rho(x')$ evaluated
at $\rho = \bar \rho$.  This covariance is the suitably scaled microscopic
truncated pair correlation \cite{DLSLD}.  For the asymmetric case discussed
here the distribution of fluctuations need no longer be given by the LDF;
in fact, $\delta^2 {\cal F}/\delta \rho (x) \delta \rho(x')$ is
discontinuous at $\bar \rho = 1/2$ in the interior of region $C$ of the
phase diagram, i.e., where $\rho_a>1/2>\rho_b$.  Furthermore, the
fluctuations there are no longer Gaussian.

   To see this discontinuity in an explicit example let
$\rho(x) = {1 \over 2} + \epsilon \Theta(x-s)$, with
$\rho_b \leq {1 \over 2} - \epsilon \leq \rho_a$; here it is easy to
compute $\tilde{\cal F}(\epsilon)\equiv{\cal F}(\{\rho(x)\})$.  First,
$F_\rho(x) = 1-\rho(x)$ if $\epsilon > 0$ and
$F_\rho(x)={1\over2}-\epsilon(1-s)$ if $\epsilon<0$; the constancy of
$F_\rho$ for $\epsilon < 0$ is due to the concave envelope construction
(\ref{CE}).  Then from (\ref{result1}) (we give only the small-$\epsilon$
behavior):
 \begin{equation}
\label{Feps}
\tilde{\cal F}(\epsilon) \simeq \left\{
        \begin{array}{ll}
       4(1-s)\epsilon^2+\ldots\;, 
          &\hbox{\hskip10pt if $\epsilon>0$,}\\
       4(1-s)\left(1-{s\over2}\right)\epsilon^2+\ldots\;, 
          &\hbox{\hskip10pt if $\epsilon<0$.}
        \end{array}
\right.
 \end{equation}
 The discontinuity of
$\partial^2\tilde{\cal F}(\epsilon)/\partial\epsilon^2$ at $\epsilon=0$
signals that the fluctuations are anomalous (non-Gaussian).  (Note that for
$s=0$, $\partial^2\tilde{\cal F}(\epsilon)/\partial\epsilon^2$ is
continuous at $\epsilon=0$ and is in fact equal to the inverse of the
variance in the total number of particles \cite{DLSALD}.)

  These non-Gaussian fluctuations can be observed by considering the total
number $N_s$ of particles on lattice sites in $[s,1]$, i.e., sites $i$ with
$sL\leq i\leq L$.  A calculation using the results of \cite{Mal} for the
microscopic probabilities shows that $[N_s-(1-s)L/2]/\sqrt L$ converges, as
$L\to\infty$, to a random variable $\mu$ with a well defined but
nonsymmetric (and non-Gaussian) distribution having density
 \begin{equation}
 p(\mu)={8\over\pi s^{3/2}(1-s)^2}\int_0^\infty dt\,t^2
   e^{-\bigl[{t^2\over s}+{2(2\mu^2+2\mu t+ t^2)\over 1-s}\bigr]}.
\label{pmu}
 \end{equation}
 This is in contrast with equilibrium systems (not at a phase transition)
for which statistical mechanics \cite{Georgii} predicts Gaussian
fluctuations of the number of particles in a macroscopic region (the
variance being related to the compressibility).  For large values of
$|\mu|$, (\ref{pmu}) yields
 \begin{equation}
 \label{p_asymp}
 -\log p(\mu) \simeq \left\{
        \begin{array}{ll}
       \displaystyle{4\mu^2\over1-s}\;, 
          &\hbox{\hskip10pt if $\mu\gg1$,}\\
  \noalign{\vskip2pt}
       \displaystyle{4\mu^2\over(1-s)(1+s)}\;, 
          &\hbox{\hskip10pt if $\mu\ll-1$.}
        \end{array}
\right.
 \end{equation}
 This agrees with the results of a large deviation calculation
\cite{DLSALD} of the probability $\exp[-L\hat{\cal F}(\epsilon)]$ of
observing mean density ${1\over2}+\epsilon$ in the interval $[s,1]$, with
no other constraints: the small-$\epsilon$ behavior of
$\hat{\cal F}(\epsilon)$ agrees with the large-$\mu$ asymptotics of
(\ref{p_asymp}) under the identification $\mu=\sqrt{L}(1-s)\epsilon$.
(Note that (\ref{p_asymp}) differs from (\ref{Feps}) because for
(\ref{Feps}) a constraint is imposed also in the region $[0,s]$.)

 {\it Derivation:} To obtain $\cal F$ in (\ref{result1}) and
(\ref{result2}) we use the exact expression for the measure $P_L(\tau)$
provided by the matrix method \cite{DEHP}.  However, rather than
calculating the probability of a given macroscopic profile $\rho(x)$
directly by summing $P_L(\tau)$ over all configurations $\tau$
corresponding to that profile, as we did for the symmetric case in
\cite{DLSLD}, we follow here a different path, which has its origin in an
{\it a posteriori} observation made in \cite{DLSLD}.  We noted there that
while ${\cal F}_s(\{\rho(x)\};\rho_a,\rho_b)$ is nonlocal, it possesses a
certain ``additivity'' property which, if it could have been established
independently, would have yielded ${\cal F}_s$.  This is exactly what we do
for the ASEP: we first derive an additivity property from the matrix
representation of $P_L(\tau)$, then use the additivity to obtain $\cal F$.
Full details are given in \cite{DLSALD}; here we give only a partial sketch
of the arguments.

Let us introduce the LDF
 ${\cal F}_{[a,b]}(\{\rho(x)\};\rho_a,\rho_b)\simeq L^{-1}
 \log P_{L(b-a)}(\{\rho(x)\};\rho_a,\rho_b)$
 for a system of $L(b-a)$ lattice sites in contact with reservoirs at
densities $\rho_a$ and $\rho_b$.  Let $K(\rho_a,\rho_b)$ be as in
(\ref{Kdef}) and define
 \begin{eqnarray}
  {\cal H}_{[a,b]}(\{\rho(x)\};\rho_a,\rho_b)&& \\
  &&\hskip-40pt \equiv\;\nonumber
  {\cal F}_{[a,b]}(\{\rho(x)\};\rho_a,\rho_b) + (b-a)K(\rho_a,\rho_b).
 \end{eqnarray}
  It is shown in \cite{DLSALD} that for $\rho_a>\rho_b$ we have
 \begin{eqnarray}
\label{additivity-asym1}
{\cal H}_{[a,b]} (\{\rho(x)\};\rho_a,\rho_b) 
  &=&  \sup_{ \rho_b \leq \rho_c \leq \rho_a}
    \Bigl[{\cal H}_{[a,c]} (\{\rho(x)\};\rho_a,\rho_c) \nonumber \\
  &&\hskip10pt +\; {\cal H}_{[c,b]} (\{\rho(x)\};\rho_c,\rho_b)\Bigr] .
 \end{eqnarray}
 while for $\rho_a<\rho_b$, 
 \begin{eqnarray}
\label{additivity-asym2}
 {\cal H}_{[a,b]} (\{\rho(x)\};\rho_a,\rho_b) \hskip-81pt \\
 &=&  \min\left[
    {\cal H}_{[a,c]} (\{\rho(x)\};\rho_a,\rho_a) +
    {\cal H}_{[c,b]} (\{\rho(x)\};\rho_a,\rho_b), \right. 
\nonumber \\ 
 &&\hskip20pt\left. {\cal H}_{[a,c]} (\{\rho(x)\};\rho_a,\rho_b) +
   {\cal H}_{[c,b]} (\{\rho(x)\};\rho_b,\rho_b) \right]. \nonumber
\end{eqnarray}
 Equations (\ref{additivity-asym1}) and (\ref{additivity-asym2}) are the
additivity relations for the ASEP.  

 In this letter we will sketch a derivation of (\ref{additivity-asym1}) 
in the special case $q=0$; for the general case and for the derivation of 
(\ref{additivity-asym2}) see \cite{DLSALD}.  We use the 
matrix formula \cite{DEHP} for a system of $N$ sites:
 \begin{equation}
 \label{exact_matrix}
 P_N(\tau) 
  = {\langle \rho_a|\Pi_{i=1}^N (D \tau_i + E(1 - \tau_i) | \rho_b \rangle
  \over \langle \rho_a|(D + E)^N| \rho_b \rangle},
 \end{equation}
 where the operators $D, E$ and vectors $\langle\rho|,|\rho\rangle$ 
satisfy 
 \begin{eqnarray}
 \label{algebra}
 DE &=& D + E,  \\
 \label{eigen}
 \langle \rho | E = {1 \over \rho } \langle \rho|, &\qquad&
 D | \rho \rangle = {1 \over  1 -\rho } | \rho \rangle.
 \end{eqnarray}
 If (\ref{eigen}) is extended to complex values of $\rho$ we may write the
exact additivity formula
 \begin{eqnarray} 
 \label{exact}
  {\langle \rho_a | X_0 X_1 | \rho_b  \rangle 
     \over \langle \rho_a |  \rho_b \rangle } && \\
 &&\hskip-40pt\nonumber
   =\; {1 \over 2 \pi i} \oint { (\rho_a - \rho_b) d \rho 
  \over (\rho_a - \rho) ( \rho - \rho_b)}
  {\langle \rho_a |X_0| \rho \rangle \over \langle \rho_a | \rho \rangle }
  {\langle \rho |X_1| \rho_b \rangle \over \langle \rho | \rho_b \rangle },
 \end{eqnarray}
 where $X_0,X_1$ are arbitrary polynomials in $D$ and $E$ and the contour is
a circle $|\rho|=R$ with $\rho_b < R < \rho_a$.  To obtain (\ref{exact}),
 note that it suffices to take $X_i=E^{p_i}D^{q_i}$,
since from (\ref{algebra}) any polynomial in $D$ and $E$ can be written as a
sum of such terms. The cases $q_0=0$ 
  or  $p_1=0$ are immediately obtained
from (\ref{eigen}) and the residue calculus; the general case
  follows by an inductive argument, as the case $q_0,p_1$ 
 can be reduced to
the cases $q_0-1,p_1$ and $q_0,p_1-1$ using (\ref{algebra}) on the left hand
side of (\ref{exact}) and the corresponding identity $(1-\rho)^{-1}\rho^{-1}=
(1-\rho)^{-1}+\rho^{-1}$ on the right hand side. 
 Now
the weights
 $\langle \rho_a |X_0| \rho \rangle/\langle \rho_a | \rho \rangle$ and
 $\langle \rho |X_1| \rho_b \rangle/\langle \rho | \rho_b \rangle$ in
(\ref{exact}) are polynomials in $1/(1-\rho)$ and $1/\rho$, respectively, with
positive coefficients, so that the integrand has a Laurent series for
$\rho_b<|\rho|<\rho_a$ with positive coefficients and hence is a convex
function of real $\rho$ for $\rho_b<\rho<\rho_a$; the minimum
 $\rho_{\rm min}$ for such real $\rho$ (which must occur since there are
poles at $\rho=\rho_a$ and $\rho=\rho_b$) is a saddle point for the complex
integral.  Thus, if $\rho(x)$ is a given profile and 
$X_0$ (respectively $X_1$) is a sum of products of
$L(c-a)$ (respectively $L(b-c)$) factors of $D$ or $E$ consistent with the
left (respectively right) part of this profile, and we assume that
that the weights depend exponentially on $L$, we expect the integral to be
dominated by this saddle point, leading to
 \begin{eqnarray} 
 \label{asymp}
  {1\over L}\log{\langle \rho_a | X_0 X_1 | \rho_b  \rangle 
     \over \langle \rho_a |  \rho_b \rangle } && \\
 &&\hskip-70pt\nonumber
   \simeq
   \inf_{\rho_b\le\rho\le\rho_a}
   \left\{{1\over L}\log{\langle \rho_a |X_0| \rho \rangle \over 
            \langle \rho_a | \rho \rangle }+
  {1\over L}\log{\langle \rho |X_1| \rho_b \rangle \over
 \langle \rho| \rho_b \rangle }\right\}.
 \end{eqnarray}
 Using (\ref{asymp}), (\ref{exact_matrix}), and the relation
 \begin{equation}
N^{-1}\log{\langle\rho_a|(D + E)^N|\rho_b\rangle\over
   \langle\rho_a|\rho_b\rangle}\simeq -K(\rho_a,\rho_b),
 \end{equation}
 we obtain (\ref{additivity-asym1}).  Thus the rather surprising supremum
in (\ref{additivity-asym1}), which corresponds to a choice of the least
probable alternative, arises mathematically through the contour
representation (\ref{exact}) and its domination by a real saddle point; we
still lack a physical explanation of (\ref{asymp}).

 To go from (\ref{additivity-asym1}) to (\ref{result1}) we divide our system
into $n$ parts of equal length, the $k^{\rm th}$ interval being
$[(k-1)/n,k/n]$, $k=1,\ldots,n$.  Now note that for very large $n$ most of
these intervals must have reservoir densities $\rho_{k-1},\rho_k$ which are
nearly equal, and that the LDF for these intervals is approximately given by
(\ref{Feq}) with $r\simeq\rho_{k-1}\simeq\rho_k$.  On the other hand, the
total length of the intervals for which this is not true (corresponding to
points of discontinuity in the function $F_\rho$) will approach $0$ as
$n\to\infty$; these considerations lead then directly to (\ref{result1}) in
this limit.  We pass from (\ref{additivity-asym2}) to (\ref{result2}) by a
similar process of subdivision, but the argument is even simpler since each
interval except one has equal reservoir densities. 

{\it Conclusion:} It would be interesting to know how key results for the
our simple model---the additivity
(\ref{additivity-asym1},\ref{additivity-asym2}) for the LD functions, the
suppression or enhancement of deviations of the density profile from its
typical value as the reservoirs and the internal field cooperate or
compete, and the non-Gaussian fluctuation (\ref{pmu}) of the number of
particles in a box of length $cL$, $0<c<1$---extend to more realistic
systems, and whether they could be understood by a dynamical approach, as
is done for the symmetric case in \cite{BDGJL}

\acknowledgments
 We thank T.~Bodineau, G.~Giacomin, and J.~M. Ortiz de Z\'arate for helpful
discussions.  The work of J.~L.~Lebowitz was supported by NSF Grant
DMR--9813268, AFOSR Grant F49620/0154, DIMACS and its supporting agencies,
and NATO Grant PST.CLG.976552.  J.L.L.  acknowledges the hospitality of the
Institut Henri Poincar\'e, where a part of this work was done.

\end{document}